
	     \magnification=1200
	     \hsize    = 165 true mm 
	     \vsize    = 220 true mm  
	     \parskip  = 3 true  pt plus 1 true pt minus 1 true  pt

	\def\centreline{\centerline}


	\def\dal{\displaystyle{{\hbox to 0pt{$\sqcup$\hss}}\sqcap}}
	\centreline{\bf Positive Energy for Asymptotically Anti-de
	Sitter Spaces}
	\vskip 0.2 true cm
	\centreline{E. Woolgar}
	\centreline{Dept. of Mathematics, Statistics, and Computer Science}
	\centreline{Dalhousie University, Halifax, Nova Scotia, Canada B3H 3J5}
	\centreline{\it and}
	\centreline{Dept. of Mathematics and Statistics, University of New Brunswick}
	\centreline{Fredericton, New Brunswick, Canada E3B 3A5}
	\vskip 0.35 true cm
	\noindent
	In 1966, Roger Penrose$^1$
	suggested that an appropriate quasi-local measure of the
	energy density content of a region of space would be the focusing power of the
	region on a passing null geodesic congruence. This {\it geometric lens effect}
	would be sensitive to both non-gravitational and gravitational energy because
	it depends on both the Ricci and the Weyl curvature in the region;
	the former governs the converging power of the lens and the latter the
	astigmatism, which for a geometric lens produces convergence in the congruence
	at second order. If the Ricci curvature has
	no negative timelike components,
	the geometric lens will have positive convergence
	(provided the region of space contains
	any ``generic'' curvature at all). This will be the case whenever the Einstein
	equation holds, provided a suitable {\it energy condition} is imposed on the
	stress-energy tensor describing the matter.

	At about the same time, Shapiro$^2$ noted that a gravitating
	positive mass would cause a {\it time-delay effect} on null geodesics passing
	near to the source, and he proposed a test of relativity (today one of the
	most accurate tests$^3$) based on measuring the time delay caused by the
	Sun's gravitational field. Conversely, in
	the presence of a negative mass source, neighbouring
	null geodesics are {\it time-advanced} with respect to those farther away,
	and this is incommensurate with the positive convergence of null geodesics
	induced by the geometric lens effect. This contradiction was used by Penrose,
	Sorkin, and the present author to prove a positive energy theorem
	for asymptotically flat spacetimes$^4$.
	The proof essentially ensures the stability
	of flat spacetime against certain classical and semi-classical decay processes
	because it shows that no suitable final state of lower energy exists.

	In this note, an argument is constructed for the case of asymptotically
	Anti-de Sitter (AdS) spacetimes$^5$.
	Attention will be restricted to those spacetimes whose first-order
	deviation from the AdS background is that of the AdS-Schwarzschild solution.
	This restriction is not as strong as the similar one for asymptotically flat
	spacetimes, wherein it would preclude gravitational radiation, because
	gravitational radiation is not present near infinity
	in any asymptotically AdS spacetime.
	The basic approach is very similar to that used in the asymptotically flat
case.
	One searches for {\it Infinite
	Achronal Null Geodesics},
	and these will exist whenever the mass is not positive.
	However, when a converging geometric lens is present (which is not the case in
	exact AdS spacetime, but is the case generically --- provided a suitable
	energy condition holds), such objects cannot
	exist. The reason is that any null geodesic passing through such a lens will
	develop conjugate points --- it will {\it focus} with a neighbouring
	geodesic --- and it must do so within finite affine distance. Whenever a
	conjugate point exists, there is always a {\it cut point} at or before it,
	this being a point at or immediately beyond which
	the null geodesic encounters timelike curves
	eminating from an earlier point along the geodesic. Beyond the cut point,
	the null geodesic cannot be achronal. Thus, an infinite achronal null geodesic
	would be an infinite null geodesic without conjugate points, which is
	impossible in generic asymptotically AdS spacetimes on which a suitable energy
	condition holds. The energy condition refers {\it only to the tracefree part}
	of the energy, not to the cosmological constant, which is negative here. A
	sufficient energy condition for present purposes is the Borde condition$^6$
	stated
	in the appendix, which is clearly insensitive to the cosmological constant.
	This condition or a related one might even hold for quantum matter and
	gravitons, in which case this positivity proof should generalise to
	{\it semi-classical} gravity.
	Also, it is remarkable that the proof herein
	relies only on an extraordinarily weak causality condition; see below.

	To find an infinite
	achronal null geodesic, first consider the metric for AdS spacetime
	$$\eqalign{ds^2=& \sec^2\xi\big ( -dt^2 + d\xi^2 +\sin^2\xi d\Omega^2 \big
)\cr
	=& \sec^2\xi\ ds^2_0\quad ,\cr}\eqno{(1)}$$
	where $d\Omega^2$ is the standard 2-sphere metric. This form may appear
alittle
	less familiar than other forms, but it will be useful in what follows.
	While AdS spacetime is a solution of the vacuum Einstein
	equations {\it with} negative cosmological constant, the cosmological constant
	does not appear in the metric in the above form. AdS spacetime is
	often said to contain Closed Timelike Curves. These are produced by
	periodically identifying $t$ in the metric, but
	the Einstein equations do not demand such an identification, and it will {\it
	not} be made here.
	This is sometimes called Universal AdS spacetime. If there exists an Infinite
	Achronal Null Geodesic in such
	a {\it universal covering spacetime}, then this is an infinite
	null geodesic without conjugate points, and it
	will project to an infinite null geodesic
	in the quotient space which will also have no conjugate points. Thus,
	it suffices to
	investigate the universal covering space.
	Hence, we have $t\in (-\infty,\infty)$;
	we also have $\xi\in[0,{{\pi}\over 2})$.

	The topology here is $R^4$, but
	$ds^2_0$ is the ``natural'' metric on
	$R\times S^3$. Thus, AdS spacetime is
	conformeomorphic to a part of the Einstein cylinder, the ``half-cylinder''
	$\xi< \pi/2$. The conformal factor is $\cos \xi$.
	Therefore, the topology (and conformal geometry) may be said to be
	that of a 3-hemisphere (say the Southern Hemisphere of some 3-sphere)
	$\times R$, the $R$ factor being time.
	The equator, defined by $\xi={{\pi}\over
	2}$, is a 2-sphere and is a
	boundary-at-infinity (${\cal I}$, called {\it scri}),
	since the conformal factor $\cos \xi$ is zero there.
	Now, because the conformal metric $ds^2_0$ has the product form,
	null geodesics of the AdS spacetime
	project to great circles on the hemisphere. In order that $ds^2_0=
	-dt^2+d\Omega^2_3=0$,
	for $d\Omega^2_3$ the 3-sphere metric, these great circles must be traversed
	with unit speed; {\it i.e.}, unit arc length along the great circles must be
	traversed in unit $t$. This prescription solves the null geodesic equations in
	AdS spacetime. Recall that great circles on spheres converge at two
	antipodal points. This means that the great circles that leave some point
	with coordinates $(\xi={{\pi}\over 2},\theta,\phi)$ on the equator will all
	reconverge at the antipodal point,
	also on the equator, whose coordinates are $(\xi={{\pi}\over
2},\pi-\theta,\phi
	\pm\pi)$.
	Furthermore, since all the great circles from any point to its antipodal point
	have equal arc
	length ($\pi$), if all the null geodesics are emitted simultaneously,
	they will reconverge simultaneously as well.
	A circle's-worth of these geodesics traverses the equator itself (in
	one less dimension, there would be a pair --- an $S^0$'s-worth --- of these,
	one going east-to-west and one going west-to-east). Many more go ``through
	spacetime''; {\it i.e.} they leave the equator. Some of these remain always
	close to the equator, while some go near to the South Pole, and more lie
	somewhere in between.

	Recall the AdS-Schwarzschild metric, which is an exact solution
	of the vacuum equations with cosmological constant. The metric can be given
	in the form
	$$ds^2=-\bigg ( 1- {{\Lambda}\over 3}r^2 - {{2M}\over r} \bigg ) dt^2
	+{{dr^2}\over {\big ( 1-{{\Lambda}\over 3}r^2-{{2M}\over r}\big ) }}
	+ r^2 d\Omega^2\quad ,
	\eqno{(2)}$$
	with $\Lambda<0$ being the cosmological constant. A rescaling
	$(s,t,r)\longrightarrow {\sqrt {{-3}\over {\Lambda}}}(s,t,r)$
	removes the cosmological constant
	and one obtains
	$$\eqalignno{
	ds^2=&-\bigg ( 1+r^2-{{2m}\over r}\bigg )dt^2 + {{dr^2}\over {\big (
1+r^2-{{2m}
	\over r}\big ) }} + r^2 d\Omega^2\quad ,&{\rm (3a)}\cr
	=&\sec^2 \xi \biggl [ ds^2_0 +{{2m\cos^3\xi}\over {\sin \xi}} \bigg ( dt^2
	+{{d\xi^2}\over {\big ( 1-{{2m\cos^3\xi}\over {\sin\xi}}\big ) }} \bigg )
	\biggl ]\quad ,
	&{\rm (3b)}\cr
	=&\sec^2\xi d\sigma^2\quad ,&{\rm (3c)}\cr}$$
	where $m={\sqrt{{-3}\over {\Lambda}}}M$, $\xi\in [0,{{\pi}\over 2})$, and $r
	=\tan \xi$.
	Consider an arbitrary spacetime whose universal covering space $({\cal
U},ds^2)$
	has topology which, outside of a cylinder,
	is $R \times S^2\times (b,\infty)$. Call this region
	${\cal U}_b$.
	The spacetime will be called {\it Asymptotically Anti-de Sitter-Schwarzschild}
	(or {\it AAdSS}) if for some $b\in R$ there exists a
	conformeomorphism of $({\cal U}_b,ds^2)$
	into the Einstein cylinder $({\cal E},ds^2_0)$ such that:
	\item{(1)}{the conformal factor admits a smooth extension which vanishes
	on the surface
	${\cal I}$ defined by $\xi={{\pi}\over 2}$, which lies in the boundary of
	the image of ${\cal U}_b$,}
	\item{(2)}{if $p,q\in {\cal I}$ are ``antipodal'' in ${\cal E}$
	in the sense that geodesics
	null in $ds^2_0$ from $p$ converge at $q$, then $\exists b$ and an $\epsilon
	>0$ such that
	the restriction of the image of $ds^2$ to the meet ${\cal S}$ of ${\cal U}_b$
	with the surface generated by these background-null curves may be given in the
	appropriate chart by}

	$$ds^2_0|_{\cal S}+2m\cos^3\xi (dt^2+d\xi^2)|_{\cal S} +
	{\cal O}(\cos^{(3+\epsilon)}\xi)
	\quad ,\eqno{(4)}$$
	\item{(3)}{any closed timelike curves that remain after passing to the
universal
	covering space are restricted to the complement of
	${\cal U}_b$.}

	\noindent
	A function $f$ is ${\cal O}(\cos^p\xi)$ if $\exists b,B \in R$ such
	that, $\forall s\in{\cal U}_b$,
	$|f(s)|<B\cos^p\xi(s)$. The fall-off condition in (4) refers to the component
	functions of the restriction of the
	image of $ds^2$ in the background $(t,\xi,\theta,\phi)$ chart.
	This specialised condition will be generalised elsewhere; for now, note that
	the metric in (3) satisfies (4). For general
	discussions of boundary conditions for asymptotically AdS spacetimes, see
	ref. 7.

	{\narrower\smallskip\noindent
	{\underbar {Theorem:}}
	If an AAdSS spacetime satisfies the Borde Energy Condition, either
	\item{(1)}{it has $M\ge 0$, or }
	\item{(2)}{it is null geodesically incomplete in $I^+({\cal I})\cap I^-
	({\cal I})$.}\smallskip}

	The proof is simple. On ${\cal E}$, consider any geodesic that is null in
	$ds^2_0$, joins
	$P\in {\cal I}$ to the ``antipodal'' point $Q\in{\cal I}$, and remains
	sufficiently near to ${\cal I}$ so as always to be
	in the image of ${\cal U}_b$, where $b$ is
	large enough so the corrections to (4) may be safely
	ignored. Such curves are always available.
	Assuming $m<0$ (hence $M<0$) , then this curve is timelike in the metric (4),
	the image of $ds^2$ on ${\cal E}$ (it is spacelike if $m>0$).
	This implies $Q\in I^+(P)$ (computed using the image of $ds^2$ and its
	extension to ${\cal I}$). Then $\exists Q'\in
	I^-(Q)\cap {\cal I}$ such that $Q'\in\partial I^+(P)$ (for otherwise, one
could
	eventually find a $Q'\in I^+(P)\cap I^-(P)\cap {\cal I}$, and neighbourhoods
of
	$Q'$ and $P$ could then be connected by closed timelike curves, contrary to
	assumption).
	By a standard theorem$^8$, from $Q'$ one can trace back along a null
	geodesic generator of $\partial I^+(P)$ which is endless except at $P$ (hence
	the geodesic never leaves $\partial I^+(P)$, so it is
	achronal). This generator is {\it not} on
	${\cal I}$, for no causal curve from $P$ which traverses ${\cal I}$ can
	reach any point in $I^-(Q)$ --- on ${\cal I}$, the extension of the conformal
	image of
	$ds^2$ agrees with $ds^2_0$, and those curves null in $ds^2_0$ only reach
	$Q$ itself, not any chronologically earlier point. Either this generator is
	incomplete or it must have infinite affine
	length in the spacetime metric\footnote{$^{\dag}$}
	{In this case, tracing back from $Q'$, either the generator
	extends from infinity, through spacetime, and back out to
	infinity, or it gets trapped, approaching a closed null generator (if one
	exists) of $\partial I^+(P)$ as its affine parameter $\to -\infty$.}
	and would be therefore an infinite achronal null geodesic, and
	would contradict Borde's focusing theorem. $\quad \dal$

	\noindent
	{\underbar {\it Remarks}:}
	For $M>0$, the fastest curves from $P$ to $Q$ are indeed
	those that never leave ${\cal I}$ --- these rule $\partial I^+(P)$ and are
	{\it not infinitely long} in the spacetime metric,
	which is not even defined on ${\cal I}$,
	so there is no contradiction.
	For exact AdS spacetime, the conditions of the Borde theorem are not satisfied
	(there is no geometric lensing), and so exact AdS space is possible, as
	it must be. This theorem does not (apparently) rule out {\it distinct} $M=0$
	AAdSS spaces --- this seems to require investigation of the ${\cal O}
	(\cos^{(3+\epsilon)}\xi)$ corrections ignored in (4). For important
	technical reasons, the above proof does {\it not} hold in the $\Lambda\to 0$
	limit; in that case, the more involved arguments of Ref. 4 are needed.

	Lastly, what about de Sitter space? Here the topology is $S^3\times R$. The
	Universe is a sort of 4-dimensional soda can, with
	future- and past-{\it timelike}
	infinities (${\cal I}^+$ and ${\cal I}^-$)
	which constitute the boundary-at-infinity of spacetime$^8$. In this
	spacetime, there are {\it event horizons} --- the past light cone of a point
	$Q\in{\cal I}^+$ does not include all of any Cauchy surface so the light cone
	itself constitutes an event horizon for the point $Q$. In de Sitter space,
	every such horizon extends back to ${\cal I}^-$, which is acceptable since
	there is no geometric lensing. However, in a generic perturbation of de Sitter
	space, such horizons cannot extend to ${\cal I}^-$, for postulate that one
did.
	This horizon is the boundary of a past set, so from every point of it there is
	a future-null generator that is endless except at $Q$. If the initial point
for
	such a generator is chosen to be $P\in{\cal I}^-$, the generator will extend
to
	$Q\in{\cal I}^+$ (unless it is incomplete), and will be an infinite achronal
	null geodesic, which cannot exist when the conditions of the Borde theorem
	hold. The implications of this are under investigation.

	\noindent
	{\underbar {\it Acknowledgements}:}
	I thank Prof. R.D. Sorkin and Prof. J. Gegenberg for various discussions and
	Dr. R. Laflamme for a question
	which led me to consider the de Sitter case. I thank the Depts. of Mathematics
	of the University of New Brunswick and Dalhousie University and specifically
	Profs. Gegenberg and B.O.J. Tupper at the
	UNB and Prof. A. Coley at Dalhousie for
	hospitality and financial support during the course of this work.

	\vskip 0.35 cm
	\centreline{\bf References}
	\vskip 0.25 true cm
	\item{(1)}{Penrose, R., in {\it Perspectives in Geometry and Relativity},
	ed. B. Hoffmann, p. 291 (Indiana University Press, Bloomington, 1966).}
	\item{(2)}{Shapiro, I.I., {\it Phys. Rev. Lett.} {\underbar{13}}, 789 (1964).}
	\item{(3)}{Will, C.M., {\it Theory and Experiment in Gravitational Physics}
	(Cambridge University Press, Cambridge, 1981).}
	\item{(4)}{Penrose, R., Sorkin, R.D., and Woolgar, E., preprint
	(1993); Penrose, R., {\it Twistor Newsletter} {\underbar {30}}, 1
	(1990); Sorkin, R.D., and Woolgar, E., {\it Proc. Fourth Can. Conf. on Gen.
	Rel. and Rel. Astrophys.}, ed. Kunstatter, G., Vincent, D.E., and Williams,
	J.G., p.206 (World Scientific, Singapore, 1992); {\it Proc. Sixth Marcel
	Grossmann Meeting on Gen. Rel.}, ed. Sato, H., and Nakamura, T., p. 754
	(World Scientific, Singapore, 1992);
	see also Ashtekar, A., and Penrose, R., {\it Twistor
	Newsletter}, (1990).}
	\item{(5)}{Gibbons, G.W., Hull, C.M., and Warner, N.P.,
	{\it Nuc. Phys.} B{\underbar{218}}, 173 (1983).}
	\item{(6)}{Borde, A., {\it Class. Quantum Gravit.} {\underbar {4}}, 343
	(1987); see also Tipler, F.J., {\it J. Diff. Eq.} {\underbar {30}}, 165
	(1978); {\it Phys. Rev.} D{\underbar {17}}, 2521 (1978); Roman, T.A.,
	{\it Phys. Rev.} D{\underbar {33}}, 3526 (1986); D{\underbar {37}}, 546
	(1988).}
	\item{(7)}{Ashtekar, A., and Magnon, A., {\it Class. Quantum
	Gravit.} {\underbar {1}}, L39 (1984); Hawking, S.W., {\it Phys. Lett.}
	B{\underbar {126}}, 175 (1983); Breitenlohner, P., and Friedman, D.Z.,
	{\it Ann. Phys.} (NY) {\underbar {144}}, 249 (1982).}
	\item{(8)}{Hawking, S.W., and Ellis, G.F.R., {\it The Large Scale Structure of
	Spacetime} (Cambridge University Press, Cambridge, 1973).}
	\vskip 0.35 true cm
	\centreline{\bf Appendix: The Borde Theorem}
	\vskip 0.25 true cm
	\noindent
	{\narrower\smallskip\noindent
	{\underbar {Borde's Focusing Theorem$^6$:}}
	Let $\gamma(t)$ be a complete affinely parametrised causal geodesic with
	tangent $l^a$ and let $l_{[a}R_{b]cd[e}l_{f]}l^cl^d\neq 0$ somewhere on
	$\gamma$. Suppose that for any $\epsilon>0$ there is a $b>0$ such
	that for any $t_1<t_2$ there is a pair of intervals $I_1<t_1$ and
	$I_2>t_2$ of lengths $\ge b$ such that
	$$\int\limits_{t'}^{t''}R_{ab}l^al^b\ dt \ge -\epsilon \quad \forall
	t'\in I_1\quad ,\quad \forall t''\in I_2\quad .$$
	Then $\gamma$ contains a pair of conjugate points. \smallskip}

	\noindent
	To obtain the result used in the text, take $l^a$ null and use the Einstein
	equations to
	replace $R_{ab}l^al^b$ by $T_{ab}l^al^b$ above. Because $l^a$ is null, the
	condition is not sensitive to the cosmological constant.
	\par\vfill\eject
	\bye